\documentclass[12pt,fleqn,twoside]{article}
\usepackage{espcrc2}
\title{
On radiative widths and a pattern of quark-diquark-gluon
configuration mixing in low-lying scalar mesons
}
\author{S.B. Gerasimov
\address
{Joint Institute for Nuclear Research,
        141980 Dubna, Moscow region, Russia}}
\begin{document}

\begin{abstract}
The earlier suggested and developed idea of quark-hadron duality,
underlying "bremsstrahlung-weighted" sum rules for total or polarized
photon interaction cross sections, is applied to the description
of excitation of light scalar mesons in gamma-gamma interactions.
The emphasis is put on the discussion of a role of the scalar diquark
cluster degrees of freedom in the radiative formation of light scalar
mesons.

\vspace{1pc}
\end{abstract}
\maketitle

\section{Introduction}
The (constituent) quark hadron duality sum rules, used here,
follow from the assumed equivalence of
two complete sets of state vectors, saturating certain integral
sum rules, one of the sets being the solution of the bound state problem
with colour-confining  interaction, while the other describes free partons.
The sum rules satisfying the assumed duality condition
have been chosen to be those related to fluctuation
of the relativistic electric dipole moment (EDM) operator in
the configuration space
of valence partons in a given system taken in the "infinite momentum" frame.
The relevance of these sum rules
has been tested in some models of quantum  field theory \cite{GeMu}
and used to derive a number of seemingly successful relations for
the hadron electromagnetic radii \cite{Ge79}
and two-photon decay widths of the lowest spin
meson resonances \cite{Ge02}.
The two-photon-meson couplings give rather direct information on the
flavour content of considered states.
Moreover, the salient
feature of our sum rule approach
enables one to put forward most distinctly the flavour content of the
states at hand and, at the same time, somehow to circumvent
many of  model-dependent
aspects of the detailed structure and dynamics of multi-component bound
quark-gluon states. Therefore they could be especially useful in the case
of hadrons with the complex and poorly understood constituent structure.
In what follows, we only mention briefly some technical details.
Varying the polarizations of
colliding photons, one can show that a linear combination of certain
$\gamma \gamma \to q\bar q $ cross-sections will dominantly collect
the $q\bar q$- states with definite spin-parity and hence
the low-mass meson resonances with the same quantum numbers.
The polarization structure of the transition matrix element
$M(J^{PC} \leftrightarrow 2\gamma)$ for the scalar meson resonance
\begin{eqnarray}
M(0^{++}\leftrightarrow 2\gamma)&=&G[(\epsilon_1 \epsilon_2)
(k_1 k_2 )\nonumber \\
&-& (\epsilon_1 k_2)(\epsilon_2 k_1)]
\end{eqnarray}
where $k_{i}^\mu$ are the momenta of photons,~$\epsilon_{i}^{\nu}$~
are polarization vectors of photons,
$G$ is a constant proportional to the coherent sum of amplitudes
describing the two-photon annihilation of partons, composing a given meson.
By definition
\begin{equation}
|G|=64\pi \Gamma_{\gamma \gamma}/m^3,
\end{equation}
where $m$ is the meson mass and $\Gamma_{\gamma \gamma}$ is
the two-photon width of a given meson.
Introducing the $\gamma \gamma $ - cross-sections $\sigma_{\bot (\|)}$
(and the integrals thereof) that refer to colliding
plane-polarized photons with the perpendicular (parallel)
polarizations, and $\sigma_{p} $  corresponding to
circularly polarized photons with parallel spins, one can then show
that the combinations of the integrals over the
bremsstrahlung-weighted and polarized
$\gamma \gamma \to q\bar q $ cross-sections,
$I_{\bot} - (1/2)I_{p},~~ I_{\|} - (1/2)I_{p},~~ I_{p}$ will be related to
low-mass meson resonances having spatial quantum numbers $J^{PC} =
0^{-+}$~ and~$ 2^{-+}$,
~~$0^{++}$~ and~$2^{++}(\lambda=0)$,
~~$2^{++}(\lambda=2)$, if we confine ourselves to the mesons with
spins $J\leq 2$ ($\lambda = 0$~~or~~$2$
being the $z$-projection of the total angular momentum of the tensor
mesons). In what  follows, we focus mainly on scalar meson sum rules
in the light quark sector.
As is known, the long-lasting experimental efforts have presently resulted in identification
of a few scalar states with masses below $2 GeV$, labelled by
isospin \cite{PDG02,Schw}:
$$I\!=\!0: f^{}_0(600)~or~ \sigma(500),
f^{}_0(980), $$
$$f^{}_0(1200 \div 1500),
f^{}_0(1506), f^{}_0(1710);$$
$$I\!=\!1/2: \kappa(800), K^*_0(1430);$$
$$I\!=\!1: a^{}_0(980), a^{}_0(1450).$$

Following \cite{Bl98,ClTo}
we assume that the listed resonances are interpreted as two meson nonets
and a scalar glueball with the mixed valence quark and
gluon configurations.
In the states lying above 1 GeV the dominant configuration is, presumably,
a conventional $q\bar{q}$ nonet mixed with the glueball of (quenched)
lattice QCD.
Below  1 GeV the states also form a nonet, where the central binding role is
played by the $SU(3)_{c(f)}$~-triplet diquark clusters
 $(qq)_{\bar{3}}(\bar{q}\bar{q})_3$ in S-wave
with some $q\bar{q}$ admixtures  in P-wave,
and maybe less important glueball part in their state vectors.
We begin with consideration of only
constituent quark and scalar diquark as the basic degrees of freedom
in the first stage of $\gamma \gamma$~-reactions:
$\gamma + \gamma \rightarrow q + \bar{q}, (qq) + (\bar{q}\bar{q})$
where the quark and diquark are treated as elementary structureless
spinor and scalar massive particles with "minimal" electromagnetic
interaction.
Evaluating cross-sections and elementary integrals
we get the sum rules for radiative widths of resonances with
$J^{PC}=0^{+ +}$
\begin{eqnarray}
\sum_{i} \frac{\Gamma(S_{i} \to 2\gamma)}{{m_{S_{i}}^3}}
\simeq \sum_{q}I_{S}(q) + \sum_{qq}I_{S}(qq),
\label{sc}
\end{eqnarray}
where
\begin{eqnarray}
I_{S}(q)= \frac{3}{16\pi^2} {\langle Q(q)^2 \rangle}^2
\frac{5\pi \alpha^2}{9m_{q}^2},
\label{scq}\\
I_{S}(qq)=\frac{3}{16\pi^2}{\langle Q(qq)^2 \rangle}^2
\frac{2\pi \alpha^2}{9m_{qq}^2}.
\label {scqq}
\end{eqnarray}

All the integrals over the parton (that is the quark and
diquark) production cross sections are rapidly converging and all
the resonance cross sections are taken in the narrow width approximation
so that for the wide scalar mesons the masses in
Eq.(\ref{sc}) have rather the meaning of the "mean value"-masses.

The term $I_{S}(qq)$ in Eqs.(\ref{sc})~and~(\ref{scqq}) corresponds
to ascribing a possible role to
scalar diquarks as a constituent triplet $(\bar d \bar s),(\bar u \bar s),
(\bar u \bar d)$ of "partons" with respective masses and electric charges
composing, at least in part, the scalar meson nonets.

Assuming now for $a_{0}(980)$ either of two limiting options:
(a)- the isovector quark-antiquark $\bar{q}q$-structure, or
(b)- the isovector diquark-antidiquark
$(\bar q \bar s)(qs)$~-configuration,
and using Eq.(\ref{sc})-(\ref{scqq}), one gets
\begin{eqnarray}
\Gamma_{\gamma \gamma}(a_{0}(980))
\simeq 1.6~(.12)~keV
\label{vqs}
\end{eqnarray}
The lower value of the width in (\ref{vqs})
is obtained if, following \cite{Pr99},
we accept for the diquark masses $m_{qs}\simeq 560$~MeV and
$m_{ud}\simeq 320$~MeV (those seem to be of minimal value as compared
to fitted mass values of many other models,
thus stressing maximally the role of the diquark configurations
 in various hadrons), while the masses of light quarks are taken
 to be $m_{u,d}\simeq 240$~MeV and $m_s\simeq 350$~MeV, according to
 \cite{Ge79}.

Both values in Eq.(\ref{vqs}) are different from
$\Gamma_{\gamma \gamma}(a_{0}(980)) = .30 \pm .10$~keV~\cite{Ams},
that is in between
two. Another evidence against the interpretation of $a_{0}(980)$~-meson
as a usual $\bar q q$-state is the quenched LQCD evaluation of
the scalar, isovector quarkonium mass
$m(0^{++},I^G=1^{-}) = 1.330(50)~GeV$~\cite{Ba02}. Therefore, it is
quite natural to suppose the $a_{0}(980)$- meson to have a mixed structure
(together with its higher-lying partner)
\begin{eqnarray}
|a_{0}(1474)\rangle = cos\theta |(1/\sqrt{2})(u\bar u - d\bar d)\rangle +
\nonumber \\
+ sin\theta |(1/\sqrt{2})((\bar d\bar s)(ds) - (\bar u\bar s)(us))\rangle, \\
|a_{0}(985)\rangle = -sin\theta |(1/\sqrt{2})(u\bar u - d\bar d)\rangle +
\nonumber \\
+ cos\theta |(1/\sqrt{2})((\bar d\bar s)(ds) - (\bar u\bar s)(us))\rangle.
\end{eqnarray}
By convention, we take here the phases of
$\sqrt{I_{S}(q)}$ and $\sqrt{I_{S}(qq)}$ as $+1$ and$-1$, respectively.
Further, with the fit $\theta \simeq 10^{o}$ to reproduce
$\Gamma_{\gamma \gamma}(a_{0}(980)) \simeq .3$~keV, one gets also
 $\Gamma_{\gamma \gamma}(a_{0}(1474)) \simeq 4.6$~keV, which should be tested
experimentally yet.
\section{A model of mixing matrices for light scalar mesons}
We turn now to a "reconstruction" of the bare masses of
two (finally) mixed scalar nonets. Taking for granted the physical masses
1474 MeV - for higher-mass, $q\bar q$-dominant isovector meson
and 985 MeV - for lower-mass $(qq) (\bar{q}\bar{q})$ one, and having defined
$\theta \simeq 10^{o}$, the diagonal and nondiagonal elements in
the $2\times 2$ mass-matrix of the isovector states are easily derived to be
$M(I=1;(q\bar q)) = 1434$~MeV, $M(I=1;((qs)(\bar{q}\bar{s})))=1005$~MeV,
and the universal, with the tentatively assumed $SU(3)$-symmetry, nondiagonal
"mass" $h=95$~MeV. In fact, it represents the transition coupling between states of
two multiplets.

The "bare" masses of the isospinor states are
$M(I=1/2;(q\bar s)) \simeq 1435$~MeV and
$M(I=1/2;((ud)(\bar{q}\bar{s}))) \simeq 812$~MeV. They correspond to
"physical" masses $m\simeq 1450$~MeV and $m \simeq 790$~MeV, according
to latest data \cite{PDG02,Schw}.

At last, to define the mass of the lightest, isoscalar "bare" state
we invoke the mass formula of the ideal-mixing-form
\begin{eqnarray}
M((ud)(\bar{u}\bar{d}))=2M((ud)(\bar{u}\bar{s}))- \nonumber \\
-M((qs)(\bar{q}\bar{s}))\simeq 620~ MeV.
\label{ideal}
\end{eqnarray}
The mixing of a glueball and $2$ pairs of
isoscalar mesons is described by the following mass matrix,
which is
diagonalized by the masses of $5$ physical states:
\begin{equation}
\label{mass}
\left(
\begin{array}{ccccc}
M_G & f & f\sqrt{2} & g & g\sqrt{2}   \\
f & M_{S_1} & 0 & h\sqrt{2} & 0   \\
f\sqrt{2} & 0 & M_{N_1} & h & h\sqrt{2}  \\
g & h\sqrt{2} & h & M_{S_2} & 0   \\
g\sqrt{2} & 0 & h\sqrt{2} & 0 & M_{N_2}  \\
\end{array}
\right)
\end{equation}
\[\hspace{1.2cm}
\Longrightarrow {\rm diag}\;(m_1,\;\!m_2,\;\!m_3\;\!m_{4},\;\!
m_{5}).
\]
$M_G$ and $M_{S_1,N_1}$ (or $M_{S_2,N_2}$)
stand for the mass of the primitive glueball, and
$S_1=s\bar{s}$ and $N_1=n\bar{n}\equiv (u\bar{u}+d\bar{d})/\sqrt{2}$
(or $S_2=((\bar n \bar s)(ns) \equiv
((\bar d \bar s)(ds)+(\bar u \bar s)(us))/\sqrt{2}$
and $N_2=(\bar u \bar d)(ud)$))~~mesons,
respectively, the subscripts $1$ or $2$ indicating
the quark (or diquark) composition of the nonet the state
belongs to; $m_i$ stand for the masses of the physical states;
$f$~~(or $g$) is the glueball--$q\bar{q}(or (\bar{qq})(qq))$--meson coupling
and $h$ is the nondiagonal quark-to-diquark pair transition coupling.
Following \cite{ClK}, we take all couplings
having dimensionality
$(mass)$, in accord with the dimensionality of the diagonal entries
of (\ref{mass}).
All quantities in (\ref{mass}) are considered to be real
numbers.

The mixing between the glueball and the low-lying
$(\bar q \bar q)(qq)$-states can be
less important also due to relative smallness
of the lowest order $gg$-to-$(\overline{qq})(qq)$ transition
amplitude as compared
to the $gg$-to-$\bar{q}q$ transition. The relevance of these arguments is
illustrated also by the (approximate) validity of mass-formulae,
Eq.(\ref{ideal}), which could be strongly violated if
the annihilation-induced mixing of different flavours would
take place. Therefore, we neglect, as a first
approximation, the coupling $g$ in the general $5\times5$ mass-matrix.
Defining the relation between the physical and bare states
\begin{equation} \left( \begin{array}{ccccc}
\underline{f_{0}(1710)}\\
\underline{f_{0}(1506)}\\
f_{0}(m_{3}) \\
\underline{f_{0}(980)} \\
f_{0}(m_{\sigma})
\end{array}\right)
=U(5)\left(\begin{array}{ccccc}
G\\
S_1\\
\underline{N_1} \\
\underline{S_2} \\
\underline{N_2},
\end{array}\right),
\label{eq:e}
\end{equation}
where masses of the underlined states are considered to be defined and
\begin{equation}
U(5)=\left(\begin{array}{ccccc}
x_1 & y_1& z_1& u_1& v_1\\
x_2& y_2& z_2& u_2& v_2\\
x_3 & y_3 & z_3& u_3& v_3 \\
x_4 & y_4 & z_4& u_4& v_4 \\
x_5 & y_5 & z_5& u_5& v_5
\end{array}\right),
\label{coef}
\end{equation}
we obtain the expression for the individual
two-photon width of a scalar meson in the form
\begin{eqnarray}
&~&\Gamma_{\gamma \gamma}(f_{0}(m_{i}))=m_{i}^{3}(125\alpha^2/7776\pi)\times
\nonumber \\
&~&\times (y_{i}A_{S_1}+z_{i}A_{N_1}+u_{i}A_{S_2}+ \nonumber \\
&~& +v_{i}A_{N_2})^2,\\
&~&A_{S_1}=\sqrt{2}/(5m_s),A_{N_1}=1/m_q, \\
&~&A_{S_2}=-\sqrt{2}/(\sqrt{5}m_{qs}),\nonumber \\
&~&A_{N_2}=-2/(5\sqrt{5}m_{ud}),
\label{width}
\end{eqnarray}
the coefficients $y_{i},..,v_{i}$ being the probability amplitudes
to find the quark configurations $S_1, N_1, S_2, N_2$ in the state
vector of the (iso)scalar meson $f_{0}(m_{i})$ with mass $m_{i}$.
The minus signs in front of $A_{N_2,S_2}$ in Eq.(\ref{width})
is the reflection of our convention about opposite signs of
the square-roots
$\sqrt{I_{S,N}(q)}$ and $\sqrt{I_{S,N}(qq)}$, defined in Eq.(\ref{scqq})
and effectively representing the fermion-quark and boson-diquark loops
in the meson-two-photon transition diagrams.
The orthogonality of the matrix $U$ in Eq.(\ref{coef}) provides
the "inclusive" sum rule to be fulfilled
\begin{eqnarray}
\sum_{i}\frac{\Gamma_{\gamma \gamma}(f_{0}(m_{i}))}{m_{i}^{3}}
= |A_{S_1}|^2+|A_{N_1}|^2+ \nonumber \\
+|A_{S_2}|^2+|A_{N_2}|^2
\end{eqnarray}
With assumed $g\simeq 0$, the unknown
elements in the mass-mixing-matrix
(\ref{mass}) are the coupling $f$ and masses $M_G$ and $M_{S_1}$, while
among the physical masses the essentially unknown is a mass
$1.2\leq m(3)\leq 1.5$~GeV~\cite{PDG02}. It seems worthwhile to mention
that starting with the evident constraint $f^2\geq 0$, we have obtained
the reasonable bounds  for these three quantities (in units of {GeV})
\begin{eqnarray}
1.31\leq m(3)\leq 1.55, 1.47\leq M_{G} \leq 1.51, \nonumber \\
1.49 \leq M_{S_1} \leq 1.69,
\label{ineq}
\end{eqnarray}
just from basic secular equations with
using the known masses of the physical mesons.
The lower-bound of $M_{G}$ and upper-bound of $M_{S_1}$ in (\ref{ineq})
are close to the values of respective masses found in \cite{ClK}
while their $m(3)=1.26$~GeV is somewhat beyond of our more general bound.
\section{Concluding remarks}
The main results of this work are the following. We applied the idea
of the $(q\bar q)-(\bar q \bar q)(qq)$-configuration-mixing to
a simpler isovector sector of two low-lying scalar nonets to fit the
two-photon width of $a_0(980)$ and extract thereby the nondiagonal element
in the mass-mixing-matrix. From the resulting more general (iso)scalar-mass
$5\times 5$-matrix, we derived the bounds on missing masses $M_G$ and
$M_{S_1}$ of the "bare" glueball and scalar strangeonium states and
poor defined mass of the so-called $f_0(1370)$-resonance.
Under the assumptions that the bare glueball-two-diquark and also
higher-lying "strangeonium"$(s\bar{s})$-two-diquark mixing can be neglected
we obtained \cite{Ge03} masses of lowest $f_0(590)$,$f_(986)$ and $f_0(1470)$~~(iso)scalar
resonances in  reasonable agreement with latest data of the E-791 and FOCUS
Collaborations \cite{Schw}. The preliminary estimates of the two-photon decay
widths can be confronted only with the experimental value $.56\pm .11$~keV
for $f_0(986)$-meson and,with stated reservations~\cite{PDG02},
for the $f_0(m_3)$-resonance, where
it is in the range of $3.8\pm 1.5 \div 5.4\pm 2.3$~~keV.
The theoretical estimates via dual sum rules \cite{Ge03} are,
roughly, two times larger than the cited data.
Clearly, for more quantitative statements we have to have new
and more accurate $\gamma\gamma$ data.
\vspace{1pc}

{\bf Acknoledgments}
\vspace{1pc}

This work was supported in part by the Bogoliubov-Infeld Foundation grant
while author visited the Theoretical Physics Department of Lodz University.
Author is grateful to Profs. M.Majewski, J. Rembielinski and W. Tybor
for helpful discussion, advices and hospitality.

\end{document}